# Heat and generalized Clausius entropy of nonextensive systems

Sumiyoshi Abe

*College of Science and Technology*, *Nihon University*,
*Funabashi*, *Chiba 274-8501*, *Japan*

Macroscopic nonextensive thermodynamics is studied without recourse to microscopic statistical mechanics. It is shown that if entropy is nonextensive, the concept of the physical temperature introduced through the generalized zeroth law of thermodynamics necessarily lead to modification of Clausius' definition of the thermodynamic entropy. It is also shown, by applying this generalized Clausius entropy to a composite nonextensive system, how the entropy and the quantity of heat behave in an arbitrary thermodynamic process. An important point emerging from this is that the entropy coefficient, which connects the microscopic and macroscopic concepts, cannot be removed from the macroscopic nonextensive theory. This fact suggests that nonextensivity may require atomism for macroscopic thermodynamics at a logical level.

There is a growing interest in thermodynamically exotic systems, which possess long-range interactions, long-term memories, (multi)fractal structures or more generically spatio-temporal complexities. A remarkable point is that, in treating such systems, it seems advantageous to relax the extensivity condition on entropy [1,2]. So far, almost all efforts devoted to understanding these nonextensive systems are based on statistical mechanical considerations. Although such a microscopic approach is essential, the state of the art is not completely satisfactory until the corresponding macroscopic thermodynamic theory is consistently formulated. Taking into account the fact that the ordinary thermodynamic framework depends crucially on the assumption of extensivity of entropy [3], it is of crucial importance to elucidate the effects of nonextensive entropy on thermodynamics.

In this paper, we study the basic structure of purely macroscopic thermodynamics of nonextensive systems *without recourse to microscopic statistical mechanical considerations*. This is done by taking the Tsallis nonextensive entropy [4] as an explicit example. The guiding principles on which we base our discussion are the generalized zeroth law of thermodynamics and the thermodynamic Legendre transform structure. A main feature of the difference between extensive and nonextensive theories is found to arise from the definition of the physical temperature through the generalized zeroth law of thermodynamics. Then, it turns out that some of ordinary thermodynamic relations including Clausius' definition of the thermodynamic entropy have to be appropriately modified. We apply these relations to a composite nonextensive system and show how the quantity of heat and the entropy consistently behave in any thermodynamic processes. We shall see that the entropy coefficient, which connects the microscopic and macroscopic concepts, is explicitly present in the macroscopic nonextensive theory.

The definition of the physical temperature should be supplied by the transitive rule, i.e. the equivalence relation, in the generalized zeroth law of thermodynamics. This is a nontrivial issue if entropy is not extensive. As a specific example, let us consider the Tsallis entropy [4]. For our purpose, it is sufficient to employ its pseudoadditivity property, which is stated as follows. If the total system is composed of two independent subsystems, $A$ and $B$, in thermal contact with each other, then the total Tsallis entropy $S_q(A, B)$ satisfies

$$S_q(A, B) = S_q(A) + S_q(B) + \frac{1-q}{k} S_q(A) S_q(B), \tag{1}$$

where $q$ is the positive entropic index whose deviation from unity describes the degree of nonextensivity of the system. This quantity is assumed to converge to the ordinary Clausius entropy in the extensive limit $q \to 1$. $k$ is the entropy coefficient having the same dimension as $S_q$. It may depend on $q$, in general, and tends to the Boltzmann constant, $k_B$, in the extensive limit. Thermal equilibrium may be characterized by the maximum of $S_q(A, B)$ under the condition of the fixed total internal energy

$$U_q(A, B) = U_q(A) + U_q(B). \tag{2}$$

From $\delta S_q(A, B) = 0$ with $\delta U_q(A, B) = 0$, we have [5-8]

$$\frac{k\beta(A)}{c(A)} = \frac{k\beta(B)}{c(B)} \equiv k\beta^*, \tag{3}$$

where

$$k\beta = \frac{\partial S_q}{\partial U_q}, \tag{4}$$

$$c \equiv 1 + \frac{1-q}{k} S_q. \tag{5}$$

Here and hereafter, we assume that $c$ is positive for any $q > 0$. $\beta^*$ is the separation constant, which characterizes the two subsystems in equilibrium. Thus, Eq. (3) defines an equivalence relation in the generalized zeroth law of thermodynamics. To identify the physical temperature, a thermometer has to be introduced and the transitive rule of the equivalence relation has to be used for the subsystems, *A* and *B*, and the thermometer. In ordinary physical situations, the thermometer is thought of as an extensive system. Hence, here is a problem regarding the composite entropy of subsystems with different values of *q*. Although such a composition law for the nonextensive entropies is not fully understood yet [9], still it is possible to formulate the generalized zeroth law of thermodynamics [10]. Consequently, the physical temperature is given by

$$T_{phys} = \frac{1}{k\beta^*} = \frac{c}{k\beta}. \tag{6}$$

The physical pressure may be defined in a similar manner by further considering mechanical equilibrium [5]. In this case, we maximize the total entropy with fixing the total volume $V(A, B) = V(A) + V(B)$. This procedure yields

$$\frac{1}{c(A)} \frac{\partial S_q(A)}{\partial V(A)} = \frac{1}{c(B)} \frac{\partial S_q(B)}{\partial V(B)} \equiv \frac{P_{phys}}{T_{phys}}. \tag{7}$$

Therefore, the physical pressure is given by

$$P_{phys} = \frac{T_{phys}}{c} \frac{\partial S_q}{\partial V}. \tag{8}$$

Before proceeding, here we make a comment on the above formulation of thermal equilibrium. The concept of "composability" [11,12] states that the total entropy of a certain kind, $S(A, B)$, is generically a function of the entropies of the subsystems, $S(A)$ and $S(B)$: $S(A, B) = f(S(A), S(B))$, where $f$ is a certain bivariate regular function. One could imagine that there exist a lot of nonextensive entropies which satisfy the composability property. Eq. (1) is nothing but an example. However, from the above discussion, it is clear that the separability condition like in Eq. (3) puts a stringent constraint on the form of the function $f(S(A), S(B))$. More explicitly, the separation as in Eq. (3) has to be realized by using the equation $k\beta(A)\,\partial f/\partial S(A) = k\beta(B)\,\partial f/\partial S(B)$. A general discussion about this point will be presented elsewhere [13].

Now, first let us consider the Legendre transform structure. Eq. (4) indicates that $\beta$ and $U_q$ form the Legendre pair of variables. This might lead to the following definition of the free energy:

$$F'_q = U_q - \frac{1}{k\beta} S_q. \tag{9}$$

This expression is however unsatisfactory since it is not written in terms of the physical temperature. The free energy should be a function of $T_{\text{phys}}$, not the unphysical variable, $(k\beta)^{-1}$. Therefore, in view of Eq. (6), we propose to define the physical free energy as follows:

$$F_q = U_q - T_{\text{phys}} \frac{k}{1-q} \ln c. \tag{10}$$

Using Eqs. (4)-(6), one can in fact ascertain that this quantity is a function of $T_{\text{phys}}$. The derivative of $F_q$ is given by

$$dF_q = dU_q - \frac{k}{1-q} \ln c \, dT_{\text{phys}} - T_{\text{phys}} \frac{dS_q}{c}. \tag{11}$$

Using the first law of thermodynamics

$$d'Q_q = dU_q + P_{\text{phys}} \, dV, \tag{12}$$

with the quantity of heat, $Q_q$, we rewrite Eq. (11) in the form

$$dF_q = d'Q_q - P_{\text{phys}} \, dV - \frac{k}{1-q} \ln c \, dT_{\text{phys}} - T_{\text{phys}} \frac{dS_q}{c}. \tag{13}$$

Thus, Clausius' definition of the thermodynamic entropy is found to be modified for nonextensive systems as follows:

$$dS_q = c \frac{d'Q_q}{T_{\text{phys}}}. \tag{14}$$

Also, the equation of state is found to be given by

$$P_{\text{phys}} = -\left(\frac{\partial F_q}{\partial V}\right)_{T_{\text{phys}}}, \tag{15}$$

which is consistent with Eq. (8).

Similarly, we can calculate the specific heat. From Eqs. (13) and (14), it follows that

$$\left(\frac{\partial F_q}{\partial T_{\text{phys}}}\right)_V = -\frac{k}{1-q}\ln c. \tag{16}$$

Using Eq. (16) in Eq. (10), we have

$$U_q = F_q - T_{\text{phys}}\left(\frac{\partial F_q}{\partial T_{\text{phys}}}\right)_V. \tag{17}$$

From this equation, the specific heat is calculated to be

$$C_{qV} = \left(\frac{\partial U_q}{\partial T_{\text{phys}}}\right)_V = -T_{\text{phys}}\left(\frac{\partial^2 F_q}{\partial T_{\text{phys}}^2}\right)_V. \tag{18}$$

Thus, in view of their standard expressions in ordinary extensive thermodynamics, both the equation of state and the specific heat are seen to remain form invariant under the present nonextensive generalization.

A main feature of nonextensivity is condensed in Eq. (14). It is of interest to apply it to a system composed of two nonextensive subsystems, $A$ and $B$, in thermal equilibrium. For this purpose, we write Eq. (14) in the form

$$\frac{d S_q(A, B)}{1+\frac{1-q}{k} S_q(A, B)} = \frac{d'Q_q(A, B)}{T_{\text{phys}}}. \tag{19}$$

Since $A$ and $B$ are in equilibrium, $T_{\text{phys}}(A) = T_{\text{phys}}(B) \equiv T_{\text{phys}}$. Let us consider an arbitrary thermodynamic process from the initial state $O$ to the final state $P$, which is composed of infinitely many isothermal and adiabatic quasistatic processes. Integrating Eq. (19), we have

$$\frac{k}{1-q} \ln\left[1 + \frac{1-q}{k} S_q(A, B)\right] - \frac{k}{1-q} \ln\left[1 + \frac{1-q}{k} S_q^{(0)}(A, B)\right]$$

$$= \int_O^P \frac{d'Q_q(A, B)}{T_{\text{phys}}}, \tag{20}$$

where $S_q(A, B)$ and $S_q^{(0)}(A, B)$ are the values of the total entropy at $P$ and $O$, respectively. For the sake of simplicity, here the initial state $O$ is taken to be the zero-temperature state. In this case, the third law of thermodynamics leads to

$$S_q^{(0)}(A, B) = S_q(A, B)\big|_{T_{\text{phys}} = 0} = 0. \tag{21}$$

Therefore, we obtain

$$\frac{k}{1-q} \ln\left[1 + \frac{1-q}{k} S_q(A, B)\right] = \int_O^P \frac{d'Q_q(A, B)}{T_{\text{phys}}}. \tag{22}$$

A point is that the rule in Eq. (1) can consistently be established if the following relation is satisfied:

$$\int_O^P \frac{d'Q_q(A, B)}{T_{\text{phys}}} = \int_O^P \frac{d'Q_q(A)}{T_{\text{phys}}} + \int_O^P \frac{d'Q_q(B)}{T_{\text{phys}}}. \tag{23}$$

To see this, let us put

$$\int_O^P \frac{d'Q_q(A)}{T_{\text{phys}}} = \frac{k}{1-q} \ln\left[1 + \frac{1-q}{k} S_q(A)\right], \tag{24}$$

$$\int_O^P \frac{d'Q_q(B)}{T_{\text{phys}}} = \frac{k}{1-q} \ln\left[1 + \frac{1-q}{k} S_q(B)\right]. \tag{25}$$

Then, from Eq. (22) with Eqs. (23)-(25), we can in fact reproduce the rule in Eq. (1). Therefore, we conclude that, for the thermodynamic process under consideration, the total entropy and the quantities of heat are connected to each other as follows:

$$S_q(A, B) = \frac{k}{1-q}\left\{\exp\left[\frac{1-q}{k}\left(\int_O^P \frac{d'Q_q(A)}{T_{\text{phys}}} + \int_O^P \frac{d'Q_q(B)}{T_{\text{phys}}}\right)\right] - 1\right\}. \tag{26}$$

Equation (23) means that the left-hand side of Eq. (22) is *extensive*. This can also be seen as follows. Let us set

$$R_q \equiv \frac{k}{1-q}\ln\left(1 + \frac{1-q}{k}S_q\right). \tag{27}$$

Using the rule in Eq. (1), we find

$$R_q(A, B) = \frac{k}{1-q}\ln\left[1 + \frac{1-q}{k}S_q(A, B)\right]$$

$$= \frac{k}{1-q}\ln\left\{\left[1 + \frac{1-q}{k}S_q(A)\right]\left[1 + \frac{1-q}{k}S_q(B)\right]\right\}$$

$$= R_q(A) + R_q(B). \tag{28}$$

This fact further leads to the conclusion that the free energy in Eq. (10) is extensive and the physical pressure in Eq. (15) is intensive, as in ordinary extensive thermodynamics. It may also be worth mentioning that $R_q$ in Eq. (27) actually corresponds to the Rényi entropy if the Tsallis entropy $S_q$ is given statistically (see Ref. [4]).

It is important to note that the entropy coefficient, $k$, which connects the microscopic and macroscopic concepts, explicitly appears and can never be removed from the macroscopic nonextensive theory. This fact suggests that although nonextensive thermodynamics developed here is purely a macroscopic theory, it may intrinsically involve atomism in itself at a logical level.

In conclusion, we have developed macroscopic nonextensive thermodynamics without recourse to microscopic statistical mechanics. This has been done using the Tsallis nonextensive entropy as an example. We have defined the physical temperature and the physical pressure through the generalized zeroth law of thermodynamics and have presented a set of nonextensive thermodynamic relations. We have shown that Clausius' definition of the thermodynamic entropy has to be appropriately modified. Then, applying the generalized Clausius entropy to a composite nonextensive system, we have seen how the nonextensive entropy and the quantity of heat behave in an arbitrary thermodynamic process. We have found that the entropy coefficient can never be removed from the macroscopic nonextensive theory.

The author would like to thank Professor A. K. Rajagopal for comments and for reading of the manuscript. This work was supported by the Grant-in-Aid for Scientific Research of Japan Society for the Promotion of Science.